\documentclass[preprint,aps,amssymb,showpacs,superscriptaddress,nofootinbib]{revtex4}
\usepackage{graphicx}
\newcommand{\del}{\partial}

\begin{document}

%\today
\preprint{}

\title{ Canonical supergravity with Barbero-Immirzi parameter}

\author{Sandipan Sengupta}
\email{sandi@imsc.res.in}
\affiliation{The Institute of Mathematical Sciences\\
CIT Campus, Chennai-600 113, INDIA.}

\author{Romesh K. Kaul} 
\email{kaul@imsc.res.in}
\affiliation{The Institute of Mathematical Sciences\\
  CIT Campus, Chennai-600 113, INDIA.}

\begin{abstract}
  A canonical formulation of the $N=1$ supergravity theory containing
  the topological Nieh-Yan term in its Lagrangian density is
  developed. The constraints are analysed without choosing any gauge.
  In the time gauge, the theory is shown to be described in terms of
  real $SU(2)$ variables.
\end{abstract}


\maketitle

%%%%%%%%%%%%%%%%%%%%%%%%%%%%%%%%%%%%%%%%%%%%%%%%%%%%%%%%%%%%%%%%%%%%%%%

\section{Introduction}
Ashtekar's canonical formulation of gravity in terms of complex
Yang-Mills connection variables has provided a gauge theoretic
interpretation of gravity\cite{ashtekar}. Subsequently, Barbero and
Immirzi have reframed this description in terms of real $SU(2)$
variables\cite{barbero}. These variables have been shown to originate
from the Holst Lagrangian density\cite{holst}, which is written in the
first order form with tetrads (e) and spin connections ($\omega$) as
independent variables :
\begin{equation} \label{LagrangianZero}
  {\cal L} ~ = ~ \frac{1}{2}e\Sigma^{\mu\nu}_{IJ}R_{\mu\nu}^{~~~ IJ}(\omega) ~ + ~
  \frac{\eta}{2}e\Sigma^{\mu\nu}_{IJ}\tilde{R}_{\mu\nu}^{~~~ IJ}(\omega)
\end{equation}
where,
\[
\Sigma_{IJ}^{\mu\nu} ~ := ~
\frac{1}{2}(e_{I}^{\mu}e_{J}^{\nu}-e_{J}^{\mu}e_{I}^{\nu})  ~,~  
R^{~~~ IJ}_{\mu\nu}(\omega) ~ := ~
\partial_{[\mu} \omega_{\nu]}^{~IJ} +
\omega_{[\mu}^{~IK}\omega_{\nu]K}^{~ ~ ~J} ~,~ \tilde{R}_{\mu\nu}^{~~~
  IJ}(\omega) ~ := ~ \frac{1}{2}\epsilon^{IJKL}R_{\mu\nu KL}(\omega).
\]
Here $\eta$, the inverse of the Barbero-Immirzi parameter, is the
coefficient of the Holst term. This additional term preserves the
classical equations of motion given by the Hilbert-Palatini action.
Thus $\eta$ appears as a free parameter in this framework.
Hamiltonian analysis of this theory based on the Lagrangian density
(\ref{LagrangianZero}) has been presented in ref.\cite{holst,sa}.

When matter is coupled to pure gravity, one needs additional terms
apart from the Holst term so that the equations of motion continue to
be independent of $\eta$. Actions containing such modifications have
been found for a few cases, e.g., spin-$\frac{1}{2}$ fermions and
$N=1, 2,4$ supergravity theories \cite{mercuri,kaul}. A superspace
formalism for $N=1$ supergravity has been presented in \cite{gates},
which reproduces the result of \cite{kaul} for this theory. It has
also been noted in ref.\cite{kaul} that although these modifications
of the Holst term for different matter couplings follow a generic
pattern in the sense that they can be written as a total divergence
after using the connection equation of motion (see \cite{mercuri}
also), they are not universal.  To emphasise, the modified Holst terms
needed to preserve the equations of motion change with the matter
content of the theory.

A universal prescription for finding a generalised action leading to a
real $SU(2)$ formulation of gravity with or without matter was proposed
in \cite{date}. This involves a Lagrangian density containing a
topological term in the form of the Nieh-Yan invariant \cite{nieh}
instead of the original Holst term:
\begin{equation} \label{ny}
{\cal L} ~ = ~ \frac{1}{2}e\Sigma^{\mu\nu}_{IJ}R_{\mu\nu}^{~~~
IJ}(\omega)+\frac{\eta}{2}I_{NY}
\end{equation}
where
\begin{equation}
I_{NY} ~ = ~
\epsilon^{\mu\nu\alpha\beta}\left[D_{\mu}(\omega)e^{I}_{\nu}\
D_{\alpha}(\omega)e_{I\beta} - \frac{1}{2}\Sigma^{IJ}_{\mu\nu}\
R_{\alpha\beta IJ}(\omega)\right] ~ ~,~ ~ D_{\mu}(\omega)e_{\nu}^I ~ :=
~ \partial_{\mu}e_{\nu}^I + \omega_{\mu~ J}^{~I}e_{\nu}^J \ .
\end{equation}
$I_{NY}$ is a total divergence, given by:
\begin{eqnarray*}
I_{NY}=\del_{\mu}(\epsilon^{\mu\nu\alpha\beta}e_{\nu}^{I} D_{\alpha}e_{\beta I})
\end{eqnarray*}
Thus it does not affect the classical equations of motion, even when
matter is coupled to the Lagrangian (\ref{ny}).

The action in (\ref{ny}) brings with it the crucial feature that
$\eta$ can be provided a topological interpretation in any theory of
gravity with or without matter. This is in contrast to the Holst
action where $\eta$ is not a coefficient of a topological term.

The Nieh-Yan term, being a topological density, can be written
uniquely in terms of the geometric variables (e, $\omega$). Thus one
does not need to look for a new `modified Holst term' whenever the
matter content changes, unlike the earlier approaches which were
matter-specific. As an elucidation of this fact, this method has been
applied to spin-$\frac{1}{2}$ fermions in ref.\cite{date}.

Here in this brief report we analyse the case of $N=1$ supergravity.
The canonical treatment of this theory has been considered earlier in
several contexts \cite{tsuda,pilati}. In ref.\cite{tsuda}, the
Hamiltonian analysis of the corresponding Holst action has been
carried out in time gauge. Here we consider a Lagrangian density
describing the same theory, but containing the Nieh-Yan invariant
instead of the Holst term in addition to the usual Hilbert-Palatini
and spin-$\frac{3}{2}$ fermionic terms. In the next section, we
exhibit the canonical formulation of this action, closely following
the analysis as given in \cite{sa,date}. Then we demonstrate that the
set of constraints in the time gauge leads to a real SU(2) description
of this theory in terms of the Barbero-Immirzi connection. We also add
a few comments on how to recover the correct transformation properties
of the fields under the action of the symmetry generators.

\section{N=1 supergravity}
The Lagrangian density for gravity coupled to spin-$\frac{3}{2}$
Majorana fermions is given by \cite{freedman}:
\begin{equation}\label{sug}
  {\cal L} ~ = ~ \frac{1}{2}e\Sigma^{\mu\nu}_{IJ}R_{\mu\nu}^{~~~
    IJ}(\omega) ~  + ~ \frac{i}{2}\epsilon^{\mu\nu\alpha\beta}
  \bar{\psi_{\mu}}\gamma_{5}\gamma_{\nu}D_{\alpha}(\omega) \psi_{\beta} 
\end{equation}
where\footnote{The Dirac matrices here obey the Clifford algebra:
  $\gamma^I\gamma^J + \gamma^J\gamma^I = 2 \eta^{IJ} ~,~ \eta^{IJ} :=
  \mathrm{diag}( -1, 1, 1, 1)$. The chiral matrix $\gamma_5 := i
  \gamma^0\gamma^1\gamma^2\gamma^3 $ and $ \sigma^{IJ} ~ ~ := ~ ~
  \frac{1}{4}~[\gamma^{I},\gamma^{J}] .$},
\begin{eqnarray*}
  D_{\mu}(\omega)\psi_{a}  ~ := ~ ~ \del_{\mu}\psi_{a} ~ + ~
  \frac{1}{2}~\omega_{\mu IJ}\sigma^{IJ}\psi_{a} ~~,~~ D_{\mu}(\omega)\bar{\psi}_{a}  ~ := ~ ~ \del_{\mu}\bar{\psi}_{a} ~ - ~
  \frac{1}{2}~\bar{\psi}_{a}~\omega_{\mu IJ}\sigma^{IJ}
  \end{eqnarray*}

  To this we add the Nieh-Yan density, to write:
\begin{eqnarray*}
  {\cal L}  ~ =~  \frac{1}{2}e\Sigma^{\mu\nu}_{IJ}R_{\mu\nu}^{~~~
    IJ}(\omega)  ~ + ~ \frac{i}{2}\epsilon^{\mu\nu\alpha\beta} \bar{\psi_{\mu}}\gamma_{5}\gamma_{\nu}D_{\alpha}(\omega) \psi_{\beta}~ + ~ \frac{\eta}{2}I_{NY}\nonumber 
\end{eqnarray*}
This can be recast as:
\begin{eqnarray}
 {\cal L} ~ =~  \frac{1}{2}e\Sigma^{\mu\nu}_{IJ}R_{\mu\nu}^{(\eta)IJ}(\omega) ~+ ~ \frac{i}{2}\epsilon^{\mu\nu\alpha\beta}
\bar{\psi_{\mu}}\gamma_{5}\gamma_{\nu}D_{\alpha}(\omega) \psi_{\beta}~+~
\frac{\eta}{2}\ \epsilon^{\mu\nu\alpha\beta}
D_{\mu}(\omega)e^{I}_{\nu}D_{\alpha}(\omega)e_{I\beta} \label{sugra} 
\end{eqnarray}
where $ ~R_{\mu\nu}^{(\eta)IJ}(\omega) ~ := ~ R_{\mu\nu}^{~~~
  IJ}(\omega)+\eta\tilde R_{\mu\nu}^{~~~ IJ}(\omega)$

The Nieh-Yan density serves as the term through which $\eta$ manifests
itself as a topological parameter in the supergravity action, and does
not show up in the classical equations of motion. This new Lagrangian
density also preserves the supersymmetry properties (on-shell) as
characterised by (\ref{sug}) since $I_{NY}$ is a total derivative.

Next we develop the analysis in the same manner as done for gravity
with spin-$\frac{1}{2}$ fermions in \cite{date}. The 3+1 decomposition
of (\ref{sugra}) can be achieved through the following parametrisation
for the tetrads and their inverses:
\begin{eqnarray*}
& & e^{I}_{t} ~ = ~ \sqrt{eN}M^{I}+N^{a}V_{a}^{I} ~ , ~ e^{I}_{a} ~ = ~
V^{I}_{a} ~ ~ ~;~ ~ ~ \nonumber \\
& & M_{I}V_{a}^{I} ~ = ~ 0 ~ , ~ M_{I}M^{I} ~ = ~ -1\nonumber \\
& & e^{t}_{I} ~ = ~ -\frac{M_{I}}{\sqrt{eN}} ~ , ~ e^{a}_{I} ~ = ~
V^{a}_{I}+\frac{N^{a}M_{I}}{\sqrt{eN}}  ~ ~; \nonumber \\
& & M^{I}V_{I}^{a} ~ = ~ 0 ~,~ V_a^I V^b_I ~ = ~ \delta_a^b ~,~ V_a^I
V^a_J ~ = ~ \delta^I_J + M^IM_J 
\end{eqnarray*} 
Also, we define $ q_{ab} ~ := ~V_{a}^{I}V_{bI}$ and $q :=
\mathrm{det}q_{ab}$ which leads to $e := det(e^{I}_{\mu}) = Nq$.

Ignoring the total spatial derivatives, the Lagrangian density can be
written as:
\begin{eqnarray*}
  & &  {\cal L} =e \Sigma^{ta}_{IJ} \partial_{t}\omega_{a}^{(\eta)IJ} ~+~
  t^{a}_{I}\partial_{t}e_{a}^{I} ~-~\bar{\pi}^{a} \del_{t}\psi_a ~-~ N H ~ - ~ N^{a}H_{a}   ~ - ~ \frac{1}{2}~\omega_{t}^{IJ}G_{IJ} ~-~ 2\bar{S}\psi_t
\end{eqnarray*} 
where $H, ~H_a,~G_{IJ}$ and $\bar{S}$ are given below in equation
(\ref{constraint}) and
\begin{eqnarray}\label{tdefn}
  & & 2e\Sigma^{ta}_{IJ} = -\sqrt{q}M_{[I}V_{J]}^{a} \nonumber\\
  & & t^{a}_{I}= \eta\epsilon^{abc}D_{b}(\omega)V_{Ic} \nonumber\\
  & & \bar{\pi}^a =  - \frac{i }{2} ~\epsilon^{abc}\bar{\psi_{b}}\gamma_5\gamma_c ~\label{momenta}
\end{eqnarray}
Here $~\bar{\pi}^a~$ is the canonically conjugate momenta associated
with $~\psi_a~$ \footnote{The functional derivative involving the
  Grassmann variables (fermions) acts on the left factor resulting in
  a sign in the definition of the conjugate momenta in
  (\ref{momenta}).}.
The last equation in (\ref{momenta}) can be inverted as:
\begin{eqnarray}\label{invert}
\bar{\psi}_a = \sqrt{q}~\bar{\pi}^{b}\gamma_a\gamma_b ~ 
\end{eqnarray}

The action does not contain the velocities associated with the gravity
fields $N,N^a, \omega_{tIJ}$ and the matter field $\psi_t$. Hence
these are Lagrange multipliers, leading to the primary constraints
$H,~H_a, G_{IJ}$ and $\bar{S}$, respectively:
\begin{eqnarray} \label{constraint}
  G_{IJ} & = & -2 D_a(\omega) \left( e \Sigma^{(\eta)ta}_{IJ}\right) -
  t^a_{[I}V_{J]}a  ~+~\bar{\pi}^a\sigma_{IJ}\psi_{a} ~\approx ~0  \nonumber\\
  H_{a} & = &e\Sigma^{tb}_{IJ}{R_{ab}^{(\eta)IJ}}(\omega) -
  V_{a}^{I}D_{b}(\omega) t^{b}_{I}~+~ \frac{i}{2}\sqrt{q}~\epsilon^{bcd}\bar{\pi}^e \gamma_b\gamma_e\gamma_5\gamma_a~D_{c}(\omega)\psi_d ~\approx ~0  \nonumber\\
  H & = & 2e^2\Sigma^{ta}_{IK}\Sigma^{tb}_{JL}\eta^{KL}R_{ab}^{(\eta)IJ}(\omega) -
  \sqrt{q}M^{I}D_{a}(\omega) t^{a}_{I}~+~\frac{iq}{2}\epsilon^{abc}\bar{\pi}^d \gamma_a \gamma_d \gamma_5 M_I \gamma^ID_{b}(\omega)\psi_c  ~\approx ~0\nonumber\\
  \bar{S} & = & D_{a}(\omega)\bar{\pi}^a~-~\frac{i\sqrt{q}}{4 \eta } \bar{\pi}^a \gamma_b \gamma_a \gamma_5 \gamma^I t_{I}^{b}  ~\approx ~0
\end{eqnarray}
where $\gamma_a$ is defined as :
\begin{eqnarray}\label{gammadefn}
\gamma_a~=~\gamma_IV_{a}^I~=~(\gamma_i-\gamma_0 \chi_i)V_{ai}
\end{eqnarray}

While $H,~H_a ,~G_{IJ}$ are the constraints for the pure gravity
sector, $\bar{S}$ is the generator of the local supersymmetric
transformations.

Following the general framework of ref.\cite{sa,date}, we introduce
the following set of convenient fields,
\begin{equation}\label{fields}
E^{a}_{i} := 2e\Sigma^{ta}_{0i} ~,~ \chi_{i} := -M_{i}/M^{0} ~,~
A^{i}_{a} ~ := ~ \omega_{a}^{(\eta)0i}-\chi_{j}\omega_{a}^{(\eta)ij} ~,~
\zeta^{i} ~ := ~ -E^{a}_{j}\omega_{a}^{(\eta)ij}
\end{equation}
alongwith the decomposition of the nine components of
$\omega^{(\eta) ij}_{a}$ in terms of three $\zeta_i$'s and six
$M_{kl}$'s ($M_{kl}=M_{lk}$) :
\begin{eqnarray}
  \omega_{a}^{(\eta)ij}=\frac{1}{2}(E_{a}^{[i}\zeta^{j]}+\epsilon^{ijk}E_{al}M^{kl})
\end{eqnarray}
In terms of the fields in (\ref{fields}), we have
$2e\Sigma^{ta}_{ij}=-E^{a}_{[i}\chi_{j]}$ and
$e\Sigma^{ta}_{IJ}\partial_{t}\omega_{a}^{(\eta)IJ}=E^{a}_{i}\partial_{t}A^{i}_{a}+\zeta^{i}\partial_{t}\chi^{i}$.
Note that the eighteen coordinate variables $\omega^{IJ}_{a}$ have
been reexpressed in terms of the twelve variables $A_{ai}$ and
$\chi_i$. The remaining six variables are the $M_{kl}$'s, whose
velocities do not appear in the Lagrangian density. Hence these are
the additional Lagrange multiplier fields.

Thus the Lagrangian density takes a simple form as follows:
\begin{eqnarray*}
  {\cal L}  &  :=  &   E^{a}_{i}\partial_{t}A^{i}_{a}~ +~
  \zeta^{i}\partial_{t}\chi^{i} ~+~ t^{a}_{I}\partial_{t}V_{a}^{I}  ~-~\bar{\pi}^{a} \del_{t}\psi_a ~-~ NH ~ - ~ N^{a}H_{a}~ - ~ \frac{1}{2}~\omega_{t}^{IJ}G_{IJ}   ~-~ 2\bar{S}\psi_t
\end{eqnarray*} 

The fields $V^{I}_{a}$ and $t^{a}_{I}$ are not really independent,
these are given in terms of the basic fields as:
$V^{I}_{a}=\upsilon^{I}_{a}$ and $t^{a}_{I}=\tau^{a}_{I}$ where
\begin{eqnarray} \label{UpsilonDefn}
\upsilon^{0}_{a} & := & -\frac{1}{\sqrt{E}}E_{a}^{i}\chi_{i} ~ ~ , ~ ~ 
\upsilon^{i}_{a} ~ := ~ \frac{1}{\sqrt{E}}E_{a}^{i}  \nonumber\\
\tau^{a}_{0} & := & \eta \epsilon^{abc} D_{b}(\omega)\upsilon_{0c} \nonumber \\
& = & \eta \sqrt{E}E^{a}_{m}\left[G^{m}_{\mathrm{rot}}-\frac{\chi_{l}}{2}\left(\frac{2f_{ml}+N_{ml}}{1 + \eta^2} + \epsilon_{mln}G^{n}_{\mathrm{boost}}\right)
-~i\bar{\pi}^b \gamma_5 (\sigma_{0m}+\frac{\chi_{l}}{2}\sigma_{ml})\psi_b\right]
 \nonumber\\
\tau^{a}_{k} & := & \eta \epsilon^{abc} D_{b}(\omega)\upsilon_{ck} \nonumber \\
& = & -\frac{\eta}{2} \sqrt{E}E^{a}_{m}\left[\frac{2f_{mk} + N_{mk}}{1 +
    \eta^2} +\epsilon_{kmn}G^{n}_{\mathrm{boost}}+~i\bar{\pi}^b \gamma_5 \sigma_{km}\psi_b\right] \label{Taudefn}
\end{eqnarray}
In the above, $f_{kl}$ and $N_{kl}$ are defined as:
\begin{eqnarray}
2f_{kl} & := & \epsilon_{ijk}E^{a}_{i}\left[( 1 + \eta^2 )
E_{b}^{l}\partial_{a}E^{b}_{j} + \chi_{j}A^{l}_{a}\right] + \eta
\left(E^{a}_{l}A_{a}^{k} - \delta ^{kl}E^{a}_{m} A^{m}_{a} -
\chi_{l}\zeta _{k}\right) + (l \leftrightarrow k) \label{FDefn}\\
N_{kl} & := & (\chi^2 - 1)( M_{kl}-M_{mm}\delta_{kl} ) + \chi_{m}\chi_{n}
M_{mn}\delta_{kl} + \chi_{l}\chi_{k} M_{mm}  - \chi_{m}(\chi_{k} M_{ml} +\chi_{l} M_{mk}) \nonumber\\
\label{NDefn}
\end{eqnarray}
We shall treat $V^{I}_{a}$ and $t^{a}_{I}$ as independent variables
and introduce associated Lagrange multipliers $\xi^{a}_{I}$ and
$\phi^{I}_{a}$ to express the equations in (\ref{Taudefn}) as
constraints.

Thus we write the full Lagrangian density as,
\begin{eqnarray}
  {\cal L} ~& = &~E^{a}_{i}\del_{t}A^{i}_{a} + \zeta^{i}\del_{t}\chi_{i}+ t^{a}_{I}\del_{t}V_{a}^{I} - \bar{\pi}^a\del_{t}\psi_a\ - NH - N^{a}H_{a} - \frac{1}{2}
  \omega_{t}^{IJ}G_{IJ}~\nonumber\\ 
& & -~\xi^{a}_{I}(V_{a}^{I}-\upsilon_{a}^{I}) -
\phi^{I}_{a}(t^{a}_{I}-\tau^{a}_{I}) ~-~ 2\bar{S}\psi_t
\end{eqnarray}

The constraints in (\ref{constraint}) can now be rewritten in terms of
the canonical fields. These can be worked out in an analogous manner
as in ref.\cite{date}. Thus the corresponding expressions for
$G^{boost}_{i}:= G_{0i}$, $G^{rot}_{i}:=
\frac{1}{2}\epsilon^{ijk}G_{jk}$, $H_a,~ H$ and $\bar{S}$ are:
\begin{eqnarray} 
  G^i_{\mathrm{boost}} & = & -\del_{a}(E^{a}_{i} ~-~ \eta
  \epsilon^{ijk}E^{a}_{j} \chi_{k}) ~+~ E^{a}_{[i}\chi_{k]}A_{a}^{k} ~+~
  (\zeta ^{i} ~-~ \chi\cdot\zeta~\chi^{i}) - t'^{a}_{[0}V_{i]a}
  \nonumber\\
  & & +~ \bar{\pi}^a \sigma_{0i} \psi_a ~ + ~ \frac{\eta}{4 M^0 E}~\epsilon^{ijk}E_{al}E^{b}_{k}\bar{\pi}^a (\gamma_j-\gamma_0 \chi_j)(\gamma_l-\gamma_0 \chi_l)(\gamma_0-\gamma_m \chi_m)\psi_b;
 \nonumber\\ 
G^{i}_{\mathrm{rot}} & = & \del_{a}(\epsilon^{ijk}E^{a}_{j} \chi_{k}
~+~ \eta E^{a}_{i}) ~+~ \epsilon^{ijk} (A^{j}_{a}E^a_k ~-~ \zeta_{j}
\chi_{k} ~-~ t'^{a}_{j}V_{a}^{k}) \nonumber\\ 
& & +~ i\bar{\pi}^a \gamma_5 \sigma_{0i} \psi_a ~ - ~  \frac{\eta}{4 M^0 E}~E_{al}\bar{\pi}^a (\gamma_{[i}-\gamma_0 \chi_{[i})E^{b}_{j]}(\gamma_l-\gamma_0 \chi_l)\gamma_j\psi_b
;
 \nonumber\\ 
H_{a} & = & E^{b}_{i}\del_{[a}A_{b]}^{i} + \zeta_{i}\del_{a}\chi_{i}
~-~ \del_{b}( t^{b}_{I}V^{I}_{a}) ~+~
t^{b}_{I}\del_{a}V^{I}_{b}\nonumber \\
& & - \frac{1}{1 + \eta^2}\left[E^{b}_{[i}\chi_{l]}A^{l}_{b} + \zeta_{i}
- \chi\cdot\zeta\chi^{i} - t^{b}_{[0}V_{i]b} - \eta \epsilon^{ijk}~
(A^{j}_{a}E^{b}_{k} + \chi_j \zeta_{k} -
t^{b}_{j}V^{k}_{b})\right]A^{i}_{a} \nonumber \\ 
& & -\frac{1}{1 + \eta^2}\left[\frac{1}{2} ~\epsilon^{ijk}\left(\eta
G^{k}_{\mathrm{boost}} + G^{k}_{\mathrm{rot}}\right) -
\chi^{i}(G^{j}_{\mathrm{boost}} - \eta G^{j}_{\mathrm{rot}})
\right]\omega_{a}^{(\eta) ij} \nonumber\\ 
& & - \frac{1}{4 (1 + \eta^2)}\frac{1}{M^0 \sqrt{E}}\epsilon^{bcd}\bar{\pi}^e \gamma_b \gamma_e (\eta-i\gamma_5)\gamma_{[a} \omega_{c]}^{(\eta) ij} (\sigma_{ij}+2\sigma_{0i}\chi_j)\psi_d \nonumber \\
& & -\frac{1}{2(1 + \eta^2)}~\frac{1}{M^0 \sqrt{E}}~\epsilon^{bcd}\bar{\pi}^e \gamma_b \gamma_e \gamma_a (\eta+ i \gamma_5) \sigma_{0i}A^{i}_{c}\psi_d   \nonumber \\
H & = &   - E^a_k \chi_k H_a + (1 - \chi\cdot\chi)\left[ E^a_i\partial_a
\zeta_i + \frac{1}{2} \zeta_i E^a_i E^b_j \partial_a E^j_b \right]
\nonumber \\
& & + \frac{1 - \chi\cdot\chi}{2 (1 + \eta^2)} \zeta_i \left[ -
G^i_{\mathrm{boost}} + \eta G^i_{\mathrm{rot}}\right] - \left( E^a_k
\chi_k V_a^I + \sqrt{q} M^I \right) \partial_b t^b_I \nonumber \\
& & - \frac{1 - \chi\cdot\chi}{1 + \eta^2} \left[ \frac{1}{2} E^a_{[i}
E^b_{j]} A^i_a A^j_b + E^a_i A^i_a \chi\cdot\zeta +
\eta\epsilon^{ijk}\zeta_i A^j_a E^a_k \right. \nonumber \\
& & \left. \hspace{3.0cm} + \frac{3}{4} (\chi\cdot\zeta)^2 - \frac{3}{4}
(\zeta\cdot\zeta) + \frac{1}{2}\zeta_i t^a_{[0}V_{i]a} - \frac{\eta}{2}
\zeta_i \epsilon^{ijk}t^a_j V^k_a \right] \nonumber \\
& & + \frac{1 - \chi\cdot\chi}{1 + \eta^2} \left[\frac{1}{\sqrt{E}}A^i_a
t^a_i + \frac{1}{2} V^i_a \left( \zeta\cdot\chi t^a_i -
\chi_i\zeta_jt^a_j + \eta \epsilon^{ijk} \zeta_j t^a_k \right) \right]
\nonumber \\
& & + \frac{1 - \chi\cdot\chi}{2(1 + \eta^2)}\left[(\eta t'^{a}_{k}~-~\epsilon^{ijk}\chi_i t'^{a}_{j})\frac{E_{al}}{\sqrt{E}}~+~2f_{kl}~+~\frac{1}{2}N_{kl}(M)~+~2 (1+\eta^2)J_{kl}\right]M^{kl} \nonumber \\
%
%& & \left. \hspace{3.0cm}~+~\frac{1}{2M^0 \sqrt{E}}~\epsilon^{lmn}E^{c}_{m}\pi_%d (\chi_n+\gamma_n \gamma_0)(1-i\eta \gamma_5)(\sigma_{0k}-\frac{1}{2}~\chi_j\s%igma_{jk})\psi_c \right]~M_{kl}  \nonumber\\
%
& & - \frac{1 - \chi.\chi}{2(1 + \eta^2)}~\frac{1}{M^0 E}\epsilon^{abc} \bar{\pi}^d \gamma_a \gamma_d \gamma_0 (\eta+i\gamma_5) \left(\sigma_{0i}A^{i}_{b}\psi_c ~+~\frac{1}{4}(\sigma_{ij}+2\sigma_{0i}\chi_j)E_{b[i}\zeta_{j]}\psi_c\right) ~ \nonumber\\
& & + ~\frac{1 - \chi\cdot\chi}{2(1 + \eta^2)}~\zeta_i ~\bar{\pi}^a (1-i\eta \gamma_5) \sigma_{0i}\psi_a \nonumber\\
\bar{S} & = & \del_a\bar{\pi}^a~-\frac{1}{1 + \eta^2}\bar{\pi}^a(1-i\eta \gamma_5)\left[\sigma_{0l}A^{l}_{a}~+~\frac{1}{2}(\sigma_{ij}+2\sigma_{0i}\chi_j )\omega^{(\eta) }_{aij}\right]-\frac{i}{4 \eta M^0 \sqrt{E}} \bar{\pi}^a \gamma_b \gamma_a \gamma_ 5 \gamma^I t_{I}^{b}\nonumber\\
\end{eqnarray}
where we have used the definitions:
%where,
%
\begin{eqnarray}
t'^{a}_{I} & := & t^{a}_{I} ~-~ \frac{\eta}{4}~\epsilon^{abc}\bar{\psi_{b}}\gamma_I \psi_{c} \nonumber\\
& = & t^{a}_{I} ~-~ \frac{\eta}{4 M^{0}\sqrt{E}}~\epsilon^{ijk}E^{a}_{k}E^{c}_{j}E^{l}_{b}~\bar{\pi}^b (\gamma_i-\gamma_0 \chi_i)(\gamma_l-\gamma_0 \chi_l)\gamma_I \psi_c \label{t'defn}\\
2J_{kl}& := &  \frac{1}{4} \epsilon^{abc}\bar{\psi}_b\gamma_k \psi_c E_{al}~~+~~(k \leftrightarrow l) \nonumber\\
& = &\frac{1}{4M^0 \sqrt{E}}~\epsilon^{iml}E_{aj}E^{b}_{m}\bar{\pi}^a(\gamma_i-\gamma_0 \chi_i)(\gamma_j-\gamma_0 \chi_j)\gamma_k \psi_b ~+~ (k \leftrightarrow l)\label{Jdefn}
\end{eqnarray}

The Hamiltonian density now reads:
\begin{eqnarray*}
{\cal H} ~=~ NH ~+~ N^{a}H_{a} ~+~ \frac{1}{2}~
\omega_{t}^{IJ}~G_{IJ} ~ +~ \xi^{a}_{I}(V_{a}^{I} ~-~ \upsilon_{a}^{I})
~+ ~\phi^{I}_{a}(t^{a}_{I}~-~\tau^{a}_{I})~ +~2\bar{S}\psi_t
\end{eqnarray*}
The constraints associated with the fields $N^{a}, N, \omega_{t}^{0i},
\omega_{t}^{ij}, \xi^{a}_{I}$ , $\phi^{I}_{a}$ and $\psi_t$
respectively are:
\begin{eqnarray*}
& & H_{a}\approx 0 ~ ~,~ ~ H\approx 0 ~ ~,~ ~
G^{i}_{\mathrm{boost}}\approx 0 ~ ~,~ ~ G^{i}_{\mathrm{rot}}\approx 0
\\ \label{Constraint1}
& & V_{a}^{I} - \upsilon_{a}^{I}\approx 0 ~ ~,~ ~ t^{a}_{I}-\tau^{a}_{I}
\approx 0~~,~ ~ \bar{S} \approx 0.		\label{Constraint2}
\end{eqnarray*}
As mentioned earlier, the momenta conjugate to $M_{kl}$ are zero. The
preservation of this constraint requires:
\begin{eqnarray*}
  \frac{\delta H}{\delta M_{kl}} ~\approx~0~,
\end{eqnarray*} 
which implies:
\begin{eqnarray}\label{mconstraint}
  (\eta t'^{a}_{k} - \epsilon^{ijk}\chi_{i} t'^{a}_{j})V_{a}^{l} + f_{kl}
  + \frac{1}{2}N_{kl} + (1 + \eta^2)J_{kl} + (k \leftrightarrow l)\approx
  0
\end{eqnarray} 
where, $f_{kl}$ and $N_{kl}$ are given in (\ref{FDefn}, \ref{NDefn}).
This constraint can be solved for $M_{kl}$. Next, using $t^{a}_{I}
~\approx ~\tau^{a}_{I}$, we write
\begin{eqnarray} \label{tau}
t'^{a}_{k} & \approx &
-\frac{\eta}{2}\sqrt{E}~E^{a}_{l}\left[~\frac{2f_{kl} + N_{kl}}{1 +
\eta^2} + 2 J_{kl} + \epsilon_{kln}
G'^{\,n}_\mathrm{boost}\right]\nonumber\\
t'^{a}_{0} & \approx & \eta
\sqrt{E}~E^{a}_{l}~\left[~G'^{\,l}_\mathrm{rot} -
\frac{\chi_{k}}{2}~\left(~\frac{2f_{kl} + N_{kl}}{1 + \eta^2} + 2 J_{kl}
+ \epsilon_{kln}~G'^{\,n}_\mathrm{boost}\right)\right]
\end{eqnarray}
Using (\ref{tau}) in (\ref{mconstraint}), we obtain
\begin{equation} \label{monstraint-fermion}
2f_{kl} + N_{kl} + 2 ( 1 + \eta^2) J_{kl} \approx 0
\end{equation}
Thus the $J_{kl}$ piece captures all the contribution coming from the
spin-$\frac{3}{2}$ fermions. Note that this equation has the same form
as the one for spin-$\frac{1}{2}$ fermions \cite{date}.  This
constraint, from (\ref{tau}), further implies:
\begin{equation}\label{t'}
t'^{a}_{I} \approx 0
\end{equation}
This is exactly same as the connection equation of motion which is
obtained in the Lagrangian formulation by varying the standard
supergravity action without the Nieh-Yan term (see \cite{kaul}, for
example).

Using (\ref{t'}), the final set of constraints read:
\begin{eqnarray*} 
  G^i_{\mathrm{boost}} & = & -\del_{a}(E^{a}_{i} ~-~ \eta
  \epsilon^{ijk}E^{a}_{j} \chi_{k}) ~+~ E^{a}_{[i}\chi_{k]}A_{a}^{k} ~+~
  (\zeta ^{i} ~-~ \chi\cdot\zeta~\chi^{i})\nonumber\\
  & & +~ \bar{\pi}^a \sigma_{0i} \psi_a ~ + ~ \frac{\eta}{4 M^0 E}~\epsilon^{ijk}E_{al}E^{b}_{k}\bar{\pi}^a (\gamma_j-\gamma_0 \chi_j)(\gamma_l-\gamma_0 \chi_l)(\gamma_0-\gamma_m \chi_m)\psi_b
\label{GBoostF}\\ 
G^{i}_{\mathrm{rot}} & = & \del_{a}(\epsilon^{ijk}E^{a}_{j} \chi_{k}
~+~ \eta E^{a}_{i}) ~+~ \epsilon^{ijk} (A^{j}_{a}E^a_k ~-~ \zeta_{j}
\chi_{k} )~+~ i\bar{\pi}^a \gamma_5 \sigma_{0i} \psi_a ~\nonumber\\ 
& &  - ~  \frac{\eta}{4 M^0 E}~E_{al}\bar{\pi}^a (\gamma_{[i}-\gamma_0 \chi_{[i})E^{b}_{j]}(\gamma_l-\gamma_0 \chi_l)\gamma_j\psi_b
\label{GRotF}\\ 
H_{a} & = & E^{b}_{i}\del_{[a}A_{b]}^{i} + \zeta_{i}\del_{a}\chi_{i}
~-~ \del_{b}\left((\tau'^{b}_{i}-\chi_i\tau'^{b}_{0})\frac{E^{i}_{a}}{\sqrt{E}}\right) ~-~
\tau'^{b}_{0}\del_{a}\left(\chi_i \frac{E^{i}_{b}}{\sqrt{E}}\right)+\tau'^{b}_{i}\del_{a}\left(\frac{E^{i}_{b}}{\sqrt{E}}\right)\nonumber \\
& & - \frac{1}{1 + \eta^2}\left[E^{b}_{[i}\chi_{l]}A^{l}_{b} + \zeta_{i}
  - \chi\cdot\zeta\chi^{i} - \frac{1}{\sqrt{E}}\tau'^{b}_{[0}E_{i]b} - \eta \epsilon^{ijk}~
  (A^{j}_{a}E^{b}_{k} + \chi_j \zeta_{k} -
  \frac{1}{\sqrt{E}}\tau'^{b}_{j}E^{k}_{b})\right]A^{i}_{a} \nonumber \\ 
& & - \frac{1}{8 (1 + \eta^2)}\frac{1}{M^0 \sqrt{E}}\epsilon^{bcd}\bar{\pi}^e \gamma_b \gamma_e (\eta-i\gamma_5)\gamma_{[a} (E_{c]}^{[i}\zeta^{j]}+\epsilon^{ijm}E_{c]}^{n}M_{mn}) (\sigma_{ij}+2\sigma_{0i}\chi_j)\psi_d \nonumber \\
& & -\frac{1}{2(1+\eta^2)M^0\sqrt{E}}~\epsilon^{bcd}\bar{\pi}^e \gamma_b \gamma_e (\eta-i\gamma_5) \gamma_{a}\sigma_{0k}A_{c}^{k}\psi_d \label{DiffeoF} \\
H & = & 
(1 - \chi\cdot\chi)\left[ E^a_i\partial_a
\zeta_i + \frac{1}{2} \zeta_i E^a_i E^b_j \partial_a E^j_b \right]
~-~ \frac{1 - \chi\cdot\chi}{\sqrt{E}} \partial_b \tau'^b_0 \nonumber \\
& & - \frac{1 - \chi\cdot\chi}{1 + \eta^2} \left[ \frac{1}{2} E^a_{[i}
E^b_{j]} A^i_a A^j_b + E^a_i A^i_a \chi\cdot\zeta +
\eta\epsilon^{ijk}\zeta_i A^j_a E^a_k \right. \nonumber \\
& & \left. \hspace{3.0cm} + \frac{3}{4} (\chi\cdot\zeta)^2 - \frac{3}{4}
(\zeta\cdot\zeta) + \frac{1}{2 \sqrt{E}}\zeta_i (\tau'^{a}_{0}-\chi_k\tau'^{a}_{k})E^{i}_{a} - \frac{\eta}{2 \sqrt{E}}
\zeta_i \epsilon^{ijk} \tau'^a_j E^k_a \right] \nonumber \\
& & + \frac{1 - \chi\cdot\chi}{1 + \eta^2} \left[\frac{1}{\sqrt{E}}A^i_a
  \tau'^a_i + \frac{1}{2\sqrt{E}} E^i_a \left( \zeta\cdot\chi \tau'^a_i -
    \chi_i\zeta_j \tau'^a_j + \eta \epsilon^{ijk} \zeta_j \tau'^a_k \right) \right]\nonumber \\
& & - \frac{1 - \chi.\chi}{2(1 + \eta^2)}~\frac{1}{M^0 E}\epsilon^{abc} \bar{\pi}^d \gamma_a \gamma_d \gamma_0 (\eta+i\gamma_5) \left(\sigma_{0i}A^{i}_{b}\psi_c ~+~\frac{1}{4}(\sigma_{ij}+2\sigma_{0i}\chi_j)E_{b[i}\zeta_{j]}\psi_c\right)  \nonumber \\
& & + ~\frac{1 - \chi\cdot\chi}{2(1 + \eta^2)}~\zeta_i ~\bar{\pi}^a (1-i\eta \gamma_5) \sigma_{0i}\psi_a ~+~ \frac{1-\chi^2}{4(1 + \eta^2)} [f_{kl}+(1 + \eta^2)J_{kl}]M^{kl}
\label{HamF}\\
\bar{S} & = &\del_a\bar{\pi}^a~-\frac{1}{1 + \eta^2}~\bar{\pi}^a(1-i\eta \gamma_5)\left[\sigma_{0l}A^{l}_{a}~+~\frac{1}{4}(\sigma_{ij}+2\sigma_{0i}\chi_j )(E_{a[i}\zeta_{j]}+\epsilon_{ijl}E_{am}M^{lm})\right]
\end{eqnarray*}
where $\tau'^{a}_{I}$ is defined as
\begin{eqnarray}
  \tau'^{a}_{I} &:= & \frac{\eta}{4} \epsilon^{abc} \bar{\psi}_b\gamma_I
  \psi_c \nonumber\\
  & = & \frac{\eta}{4 M^{0}\sqrt{E}}~\epsilon^{ijk}E^{a}_{k}E^{c}_{j}E^{l}_{b}~\bar{\pi}^b (\gamma_i-\gamma_0 \chi_i)(\gamma_l-\gamma_0 \chi_l)\gamma_I \psi_c 
\end{eqnarray}\label{tau'}
and $f_{kl}$ , $J_{kl}$ and $M_{kl}$ are given by the (\ref{FDefn}),
(\ref{NDefn}), (\ref{Jdefn}) and (\ref{monstraint-fermion}).
% Time gauge
%\begin{eqnarray}
%2f_{kl} & = & (1 +
%\eta^2)\epsilon^{ijk}E^{a}_{i}E^{l}_{b}\del_{a}E^{b}_{j} +
%\eta\left(E^{a}_{k}A^{l}_{a} - \delta^{kl}E^{a}_{m}A^{m}_{a}\right) + (k
%\leftrightarrow l)\nonumber  \\
%%
%2J_{kl} & = & \frac{1}{4M^0 E}~\epsilon^{iml}E_{aj}E^{b}_{m}\bar{\pi}^a \gamma_i \gamma_j\gamma_k \psi_b ~+~ (k \leftrightarrow l)
%\end{eqnarray}
In writing $\bar{S}$, we have made use of the Fierz identity-
\begin{eqnarray*}
  \epsilon^{\mu\nu\alpha\beta}\bar{\psi}_{\mu}\gamma_{I}\psi_{\nu}\gamma^{I}\psi_{\alpha}
~=~0~~~~~~,
\end{eqnarray*}
which makes the piece proportional to $t^{a}_{I}$ dissapear.

\subsection*{Time gauge:} 
One may adopt the time gauge through the choice $~\chi_{i} = ~0~$.
Since this condition forms a second-class pair with the boost
constraint, both have to be implemented together. $G_{i}^{boost}$ can
be solved as:
\begin{equation} 
\zeta_{i} ~ = ~ \del_{a}E^{a}_{i} ~-~\bar{\pi}^a \sigma_{0i} \psi_a ~-~\frac{\eta}{4 M^0 E}~\epsilon^{ijk}E_{al}E^{b}_{k} \bar{\pi}^a \gamma_j\gamma_l\gamma_0\psi_b
\end{equation}
We can rewrite this as:
\begin{eqnarray} \label{ZetaSoln} 
\zeta_{i} ~ = ~ \del_{a} E^{a}_{i}
  ~+~\frac{1}{\sqrt{E}}{\tau'}_{0}^{a} E_{ai}~+~\frac{1}{\eta
    \sqrt{E}}\epsilon^{ijk} {\tau'}_{j}^{a} E_{ak}
\end{eqnarray}
with
\begin{eqnarray*}
  \tau'^{a}_{I}~=~\frac{\eta}{4 \sqrt{E}}\epsilon^{ijk}E_{k}^{a}E_{j}^{c}E_{bl}\bar{\pi}^b\gamma_i\gamma_l\gamma_I \psi_{c}
\end{eqnarray*}
The constraints in this gauge read:
\begin{eqnarray}\label{timegaugeconstraint}
  G^{i}_\mathrm{rot} & = & \eta~ \del_{a}E^{a}_{i} +
  \epsilon^{ijk}A_{a}^{j}E^{a}_{k}~-~\frac{1}{\eta
    \sqrt{E}}{\tau'}_{0}^{a} E_{ai}~-~ \frac{1}{\sqrt{E}}\epsilon^{ijk} {\tau'}_{j}^{a} E_{ak}
  \nonumber\\
H_{a} & = & E^{b}_{i}F_{ab}^{i} ~ -~ \frac{1}{\eta^2 \sqrt{E}}\tau'^{b}_{0}E_{bi}A_{ai}- \frac{E_{ai}}{\sqrt{E}}\left[\del_b \tau'^{b}_{i} + \frac{1}{\eta}\epsilon^{ijk}A^{j}_{b}\tau'^{b}_{k}\right] +\frac{1}{2(1+\eta^2)}\left[A^{k}_{a}-E^{i}_{a}E^{b}_{k}A^{i}_{b}\right]\zeta_{k}\nonumber \\
H & = & -\frac{\eta}{2}E^{a}_{i}E^{b}_{j} \epsilon^{ijk} \left[F_{ab}^{k}+~(\eta+\frac{1}{\eta})~ R_{ab}^{k}\right]  -\frac{1}{\sqrt{E}}\left[\del_a \tau'^{a}_{0}-\frac{1}{2\eta}(\epsilon^{ijk}E^{j}_{a}\zeta_k \tau'^{a}_{i}+E^{j}_{a}\tau'^{a}_{i}M^{ij})\right]\nonumber \\
\bar{S} & = &\del_a\bar{\pi}^a~-\frac{1}{1 + \eta^2}~\bar{\pi}^a(1-i\eta \gamma_5)\sigma_{0k}\left[A^{k}_{a}~+~\frac{1}{2}i\gamma_5(\epsilon_{jkl}\zeta_{j}+M^{kl})E_{al}\right]
\end{eqnarray}
In these equations, we have used the following definitions:
\begin{eqnarray*}
  & & \Gamma_{ai}~=~\frac{1}{2}\epsilon^{ijk}\omega_{ajk}\\
  & &  F_{ab}^{k}~=~\del_{[a}A_{b]}^{k}+\frac{1}{\eta} \epsilon^{ijk}A_{ai}A_{bj}~~,~~~
R_{ab}^{k}~=~\del_{[a}\Gamma_{b]}^{k}-\frac{1}{\eta} \epsilon^{ijk}\Gamma_{ai}\Gamma_{bj}
\end{eqnarray*}
and $\zeta^i$ is given by (\ref{ZetaSoln}). Also, in the time gauge :
\begin{eqnarray*}
M_{kl}~& =&~ (1+\eta^2)\left(\epsilon^{ijk}E_{b}^{l}\del_{a}E^{b}_{j}-\epsilon^{ijm}E_{b}^{m}\del_{a}E^{b}_{j}\delta_{kl}\right) E^{a}_{i}~+~(1+\eta^2)\left(2J_{kl}-J_{mm}\delta_{kl}\right) \\
& & ~+~\eta E^{a}_{l} A_{ak}~~+ ~~(k \leftrightarrow l)\\
2J_{kl}~& =& ~ \frac{1}{4M^0 \sqrt{E}}~\epsilon^{iml}E_{aj}E^{b}_{m}\bar{\pi}^a \gamma_i\gamma_j\gamma_k \psi_b ~~+~~(k \leftrightarrow l)
\end{eqnarray*}
Here in (\ref{timegaugeconstraint}) we have dropped terms proportional
to rotation constraints from $H_a$ and $H$.

As is evident, the dynamical variable which enters in the constraints
apart from the fermionic degrees of freedom is the Barbero-Immirzi
connection $A_a^i$. Thus in the time gauge we obtain a real SU(2)
formulation of the theory of gravity coupled to spin-$\frac{3}{2}$
fermions.

Notice that in the matter sector, $\bar{\pi}^{a}$ and $\psi_{a}$ are
not independent variables. These obey the second-class constraints:
\begin{eqnarray*}
  C^a := \bar{\pi}^{a}+ \frac{i }{2} ~\epsilon^{abc}\bar{\psi_{b}}\gamma_5\gamma_c \approx 0
\end{eqnarray*}
In order to implement these constraints, we need to go to
corresponding Dirac brackets for the matter fields $\bar{\pi}^{a},
\psi_{a}$. This then leads to the correct transformations (modulo
rotations) on the fields through their Dirac brackets with the
corresponding generators. In particular, the Dirac brackets of the
fields with the supersymmetry generator $\bar{S}$ make them transform
properly under its action.

\section{Conclusions}
We have presented a framework to incorporate the Barbero-Immirzi
parameter as a topological coupling constant in the classical theory
of $N=1$ supergravity. This is achieved through the inclusion of the
Nieh-Yan density in the Lagrangian. This additional term, being a
topological density, preserves the equations of motion and the
supersymmetry of the original action. To emphasise, this goes beyond
the earlier analysis involving the Holst action which does not allow a
topological interpretation for $\eta$.

The canonical formulation has been first developed without going to
any particular choice of gauge. This clarifies the structure of the
theory exhibiting all of its gauge freedom. In the time gauge, the
theory is shown to admit a real SU(2) formulation in terms of the
Barbero-Immirzi connection $A^{i}_{a}$.

The essential features for spin-$\frac{3}{2}$ fermions turn out to be
very similar to those for spin-$\frac{1}{2}$ fermions as described in
\cite{date}, except that here we have the additional constraint
$\bar{S}$ which acts as the generator of local supersymmetry
transformations.  The cases for $N=2, 4$ and higher supergravity
theories can be treated in exactly similar fashion. There the
constraint analysis leads to the same form of the connection equation
of motion as given here (i.e., equation (\ref{monstraint-fermion})), a
fact which is evident from the structure of the fermionic terms in
these theories. Only the expression for $J_{kl}$ in terms of the
matter fields gets modified.

The analysis here has been purely classical. However, in the quantum
theory, the presence of the topological Nieh-Yan term, which is also
CP violating, may reflect a possible non-perturbative vacuum
structure.
%It remains to be seen
%whether this new topological term in the supergravity action leads to
%any non-perturbative feature in the quantum theory.
\vspace{0.7cm}

Acknowledgements: 

We thank Prof. Ghanashyam Date and V. S. Nemani for useful discussions.

\end{document}